\documentclass[preprint,showpacs,preprintnumbers,amsmath,amssymb]{revtex4}

\usepackage{graphicx}
\usepackage{dcolumn}
\usepackage{bm}

\begin{document}

\preprint{}

\title{Photoproduction and Radiative Decay \\
of Spin 1/2 and 3/2 Pentaquarks}

\author{$^{1,2}$Xiao-Gang He\footnote{e-mail: hexg@phys.ntu.edu.tw. On leave of absence from
Department of Physics, National Taiwan University, Taipei},
$^2$Tong Li\footnote{e-mail: allongde@mail.nankai.edu.cn},
$^2$Xue-Qian Li\footnote{e-mail: lixq@nankai.edu.cn}, $^3$C.-C.
Lih\footnote{e-mail: cclih@phys.ntu.edu.tw}}
\address{$^1$Department of Physics, Peking University, Beijing\\
$^2$Department of Physics, Nankai University, Tianjin\\
$^3$Department of Physics, National Taiwan University, Taipei}


\date{\today}

\begin{abstract}
We study photoproduction and radiative decays of pentauqarks
paying particular attention to the differences between spin-1/2
and spin-3/2, positive and negative parities of pentaquarks.
Detailed study of these processes can not only give crucial
information about the spin, but also the parity of pentaquarks.
\end{abstract}

\maketitle

\section{Introduction}

Recently several experiments have reported evidences for
pentaquarks $\Theta$ and other states\cite{1,2,3}. The first
observed pentaquark state was the $\Theta(1540)$ with strangeness
$S=+1$ and was identified as a state with quark content $udud\bar
s$. This particle is an isosinglet and belongs to the
anti-decuplet multiplet in flavor $SU(3)_f$ symmetry\cite{4}.
Consequently NA49 has reported evidences for isoquated $\Xi_{3/2}$
in the anit-decuplet\cite{2}. At present there are very limited
information on the detailed properties such as the spin, the
parity and the magnetic dipole moment. Several other experiments
have also carried out searches for these particles. Some of them
reported positive and while others reported negative
results\cite{3}. One has to wait future experiments to decide
whether these pentaquark state are real. On the theoretical front,
there are also many studies trying to understand the propertices
of these possible pentaquark states\cite{5,7,8,9,10}

In this paper we explore possibilities of studying the properties
of pentaquark $\Theta$ and its partners in the SU(3) anti-decuplet
multiplet, using radiative processes involving a pentaquark $P$,
an ordinary baryon $N$ and a pseudoscalar $\Pi$. We consider two
classes of processes, the photoproduction $\gamma + N \to \Pi P$
and radiative decay $P\to N \Pi \gamma$.

In the above $N$ and $\Pi$ indicate a member in the ordinary
baryon octet and pseudoscalar octet of $SU(3)_f$, respectively.
They are given by

\begin{eqnarray}
N = (N_i^j) = \left ( \begin{array}{lll}
{\Sigma^0\over \sqrt{2}} + {\Lambda\over \sqrt{6}}&\Sigma^+&p\\
\Sigma^-&-{\Sigma^0 \over \sqrt{2}} + {\Lambda\over \sqrt{6}}&n\\
\Xi^-&\Xi^0&-{2\Lambda\over \sqrt{6}}
\end{array}
\right ),\;\;\;\; \Pi = (\Pi_i^j) = \left ( \begin{array}{lll}
{\pi^0\over \sqrt{2}} + {\eta\over \sqrt{6}}&\pi^+&K^+\\
\pi^-&-{\pi^0 \over \sqrt{2}} + {\eta\over \sqrt{6}}&K^0\\
K^-&\bar K^0&-{2\eta\over \sqrt{6}}
\end{array}
\right ).
\end{eqnarray}

$P$ is a member of the anti-decuplet ($\overline{10}$) pentaquark
multiplet. This multiplet has 10 members which can be described by
a totally symmetric tensor $P^{ijk}$ in SU(3). The 10 memebers are

\begin{eqnarray}
&&P^{111} = \Xi^{--}_{3/2},\;\;P^{112} = \Xi^-_{3/2}/\sqrt{3},
\;\;P^{122} = \Xi^0_{3/2}/\sqrt{3},\;\;P^{222} = \Xi^+_{3/2},\nonumber\\
&&P^{113} = \Sigma^-_a/\sqrt{3},\;\;P^{123} = \Sigma^0_a/\sqrt{6},
\;\;P^{223} = \Sigma^+_a/\sqrt{3},\nonumber\\
&&P^{133} =  N^0 _a/\sqrt{3},\;\;
P^{233} = N^+_a/\sqrt{3},\;\;P^{333}=\Theta^+.
\end{eqnarray}

Without $SU(3)_f$ symmetry breaking members in a $SU(3)_f$
multiplet all have the same mass. The degeneracy of mass is lifted
by the light quark mass differences, $m_u$, $m_d$ and $m_s$. Using
information on the masses of $\Theta$ and $\Xi_{3/2}$ including
the leading $SU(3)_f$ breaking effects, the masses of the
anti-decuplet members are given by\cite{5} $m_\Theta = 1542$ MeV,
$m_{\Xi_{3/2}} = 1862$ MeV, $m_{\Sigma_a} = 1755$ MeV, and
$m_{N_a} = 1648$ MeV.

Discussions for radiative processes involving a $P$, a $N$, a
$\Pi$ and a $\gamma$ with spin-1/2 pentaquarks have been carried
out in several papers\cite{5,7}. There are also some studies for
spin-3/2 pentaquarks\cite{8}, but no detailed studies of radiative
processes. In this work we will consider both spin-1/2 and
spin-3/2 cases and paying particular attention for the
differences. Since in the processes considered involve
pseudoscalar goldstone bosons $\pi$ and $K$, we will use chiral
perturbation theory to carry out the analysis.

\section{The matrix elements for radiative processes}

The leading order diagrams for the radiative processes involving a
$P$, a $N$, a $\Pi$ and a $\gamma$ are shown in Figure 1. The
electromagnetic coupling of photon with $\Pi$ and $N$ are known.
To evaluate these diagrams, we need to know the various couplings
involving pentaquarks.

\subsection{The spin-1/2 case}

There are two types of electromagnetic couplings, the electric
charge and magnetic dipole interactions. The leading chiral
electric charge and magnetic dipole couplings are given by

\begin{eqnarray}
L_e &=& \bar P i \gamma^\mu D_\mu P = \bar P_{ijk}i\gamma^\mu
(\partial_\mu
P^{ijk}-V_{\mu,l}^iP^{ljk}-V_{\mu,l}^jP^{ilk}-V_{\mu,l}^kP^{ijl}),\nonumber\\
L_{m} &=& {\mu_P\over 4} \bar P_{ijk}\sigma^{\mu\nu}
(f_{\mu\nu,l}^iP^{ljk}+f_{\mu\nu,l}^jP^{ilk}+f_{\mu\nu,l}^kP^{ijl}),
\end{eqnarray}
where $V_\mu = (1/2)(\xi^\dagger \partial_\mu \xi + \xi
\partial_\mu \xi^\dagger) + i(e/2)A_\mu(\xi^\dagger Q \xi + \xi Q\xi^\dagger)$.
Here $\xi = exp[i\Pi/\sqrt{2}f_\pi]$ and $Q=Diag(2/3,-1/3,-1/3)$
is the quark charge matrix and $A_\mu$ is the photon field.
$f_{\mu\nu,i}^j = F_{\mu\nu}(\xi^\dagger Q \xi + \xi
Q\xi^\dagger)_i^j$ with $F_{\mu\nu}$ being the photon field
strength. Expanding to the leading order, we have for each
individual pentaquark

\begin{eqnarray}
L_e&=&-eQ_i \bar P_i \gamma^\mu A_\mu P_i,\nonumber\\
L_m&=&-{e\mu_P Q_i\over 2} \bar P_i \sigma^{\mu\nu}F_{\mu\nu}P_i.
\end{eqnarray}

We note that for neutral pentuaquarks, to the leading order the
anomalous dipole moments are zero. The kappa parameter $\kappa_P =
2m_P \mu_P$ have been estimated to be of order one\cite{9}. In our
analysis we will treat it as a free parameter to see if
experimental data can provide some information.

We also need to know the strong interaction coupling of a
pentaquark with an ordinary baryon and a pseudoscalar. It can be
parameterized as

\begin{eqnarray}
&&L_{PN\Pi} = g_{PN\Pi}\bar P_{ilm} \Gamma_P \gamma^\mu (\tilde
A_\mu)^l_j N^m_k \epsilon^{ijk} + H.C.
\end{eqnarray}
In the above $\Gamma_p$ takes ``+1'' and ``$\gamma_5$'' if $P$ has
negative and positive parities, respectively. $\tilde A_\mu =
(i/2)(\xi^\dagger
\partial_\mu \xi - \xi \partial_\mu \xi^\dagger)-(e/2) A_\mu
(\xi^\dagger Q \xi - \xi Q \xi^\dagger)$.

Expanding the above effective Lagrangian to the leading order we
obtain $P-N-\Pi$ type of couplings. The results are given in Table
1.

\begin{figure}[!htb]
\begin{center}
\caption{Radiative processes involving a pentaquark, an octet
baryon and an octet meson.}
\includegraphics[width=10cm]{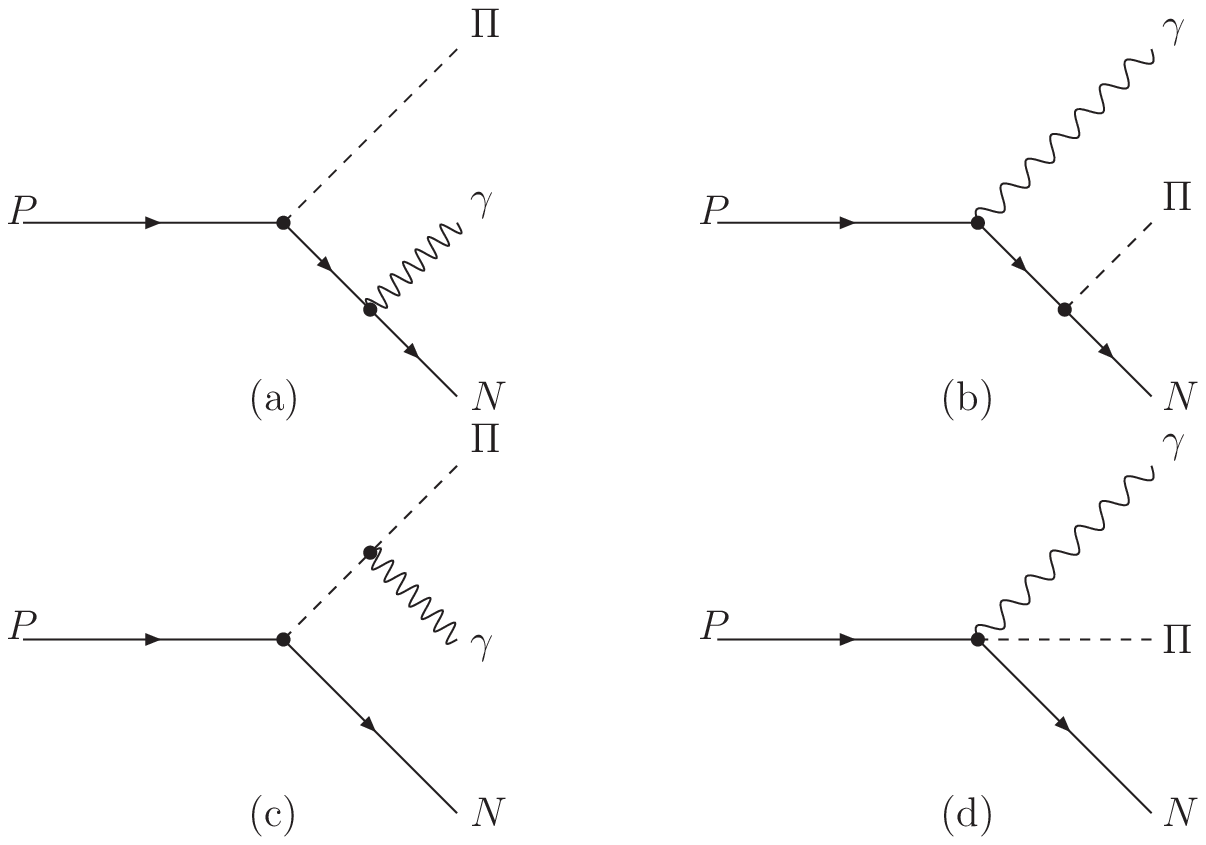}%
\end{center}
\label{fig1}
\end{figure}
\pagebreak[4]

\begin{table}[htb]
\caption{$P$-$N$-$\Pi$ couplings in unit
$g_{PN\Pi}/\sqrt{2}f_\pi$. The couplings in the tables are
understood to be in the form $-a_{PN\Pi}\bar P \Gamma_P\gamma^\mu
N \partial_\mu \Pi $. The coefficient in front of $N\Pi$ in the
second column is $-a_{PN\Pi}$.}\label{coupling-ad-b}
\begin{tabular}{|l|l|}
\hline $\Theta^+$&$-nK^+ + p K^0$\\ \hline $N^0_a$&${1\over 6}
(-3\sqrt{2} n \eta + 3\sqrt{2} \Lambda K^0 +\sqrt{6} \Sigma^0 K^0
- \sqrt{6} n \pi^0 + 2\sqrt{3} p \pi^- - 2\sqrt{3} \Sigma^-_a
K^+)$\\ \hline $N^+_a$&${1\over 6} (3\sqrt{2} p \eta -3\sqrt{2}
\Lambda K^+ + \sqrt{6} \Sigma^0 K^+ - \sqrt{6} p \pi^0 - 2\sqrt{3}
n \pi^+ + 2\sqrt{3} \Sigma^+ K^0)$\\ \hline $\Sigma^-_a$&${1\over
6} (2\sqrt{3} n K^- + 3\sqrt{2} \Lambda \pi^- + \sqrt{6} \Sigma^0
\pi^- - 3\sqrt{2} \Sigma^- \eta - \sqrt{6} \Sigma^- \pi^0 -
2\sqrt{3} \Xi^- K^0)$\\\hline $\Sigma^0_a$&${1\over 6} ( \sqrt{6}
n \bar K^0 - \sqrt{6} p K^- - 3\sqrt{2} \Lambda \pi^0 + 3\sqrt{2}
\Sigma^0 \eta - \sqrt{6} \Sigma^- \pi^+ + \sqrt{6} \Sigma^+ \pi^-
-\sqrt{6} \Xi^0 K^0 + \sqrt{6}\Xi^- K^+)$\\ \hline
$\Sigma^+_a$&${1\over 6} ( -2\sqrt{3} p \bar K^0 - 3\sqrt{2}
\Lambda \pi^+ + \sqrt{6} \Sigma^0 \pi^+ + 3\sqrt{2} \Sigma^+ \eta
- \sqrt{6} \Sigma^+ \pi^0 + 2\sqrt{3} \Xi^0 K^+)$\\ \hline
$\Xi^{--}_{3/2}$&$\Sigma^- K^- - \Xi^- \pi^-$\\ \hline
$\Xi^-_{3/2}$&${1\over 6} ( -2\sqrt{6} \Sigma^0 K^- + 2\sqrt{3}
\Sigma^- \bar K^0 -2\sqrt{3} \Xi^0 \pi^- + 2\sqrt{6} \Xi^-
\pi^0)$\\ \hline $\Xi^0_{3/2}$&${1\over 6} ( -2\sqrt{6} \Sigma^0
\bar K^0 - 2\sqrt{3} \Sigma^+ K^- + 2\sqrt{3} \Xi^0 \pi^0 +
2\sqrt{6} \Xi^- \pi^+)$\\ \hline $\Xi^+_{3/2}$&$- \Sigma^+ \bar
K^0 + \Xi^0 \pi^+$
\\\hline
\end{tabular}
\end{table}

The contact $\gamma$-$P$-$N$-$\Pi$ coupling in Figure 1.d is
obtained from a term $ie g_{PN\Pi}A_\mu \bar
P_{ilm}\Gamma_P\gamma^\mu[\Pi,Q]^l_j N^m_k \epsilon^{ijk}$
obtained by expanding $L_{PN\Pi}$.

In the following we display the matrix element for $P\to N \Pi
\gamma$. The matrix element for $\gamma N \to P \Pi$ can be
obtained by making appropriate changes of signs for the relevant
particle momenta. We have

\begin{eqnarray}
M(P \to  N \Pi \gamma) &=& {eg_{PN\Pi}\over \sqrt{2} f}
 a_{PN\Pi}\epsilon^*_\mu
 \bar N[ Q_{\Pi} \Gamma_P \gamma^\mu\nonumber\\
&-&(Q_N\gamma^\mu + {\mu_N\over 2} [\gamma^\mu, \gamma \cdot
P_\gamma]) {1\over \gamma\cdot P_\gamma +\gamma\cdot P_N - m_N}
\Gamma_P \gamma \cdot P_\pi\nonumber\\
&-&\Gamma_P \gamma \cdot P_\Pi {1\over \gamma \cdot P_N + \gamma
\cdot P_\Pi - m_P} (Q_P \gamma^\mu + {\mu_P\over 2}[\gamma^\mu,
\gamma \cdot P_\gamma])\nonumber\\
 &-& Q_\Pi {(2P_\Pi+P_\gamma)^\mu\over
(P_\Pi+P_\gamma)^2-m^2_\Pi} \Gamma_P (\gamma \cdot P_\Pi + \gamma
\cdot P_\gamma)]P.
\end{eqnarray}

For $\Theta^+ \to  n K^+ \gamma$, $a_{PN\Pi}=a_{\Theta nK} = 1$,
$Q_P=Q_\Theta = 1$, $Q_N = Q_n =0$, $Q_\Pi = Q_{K^+} = 1$. For
$\Theta^+ \to  p K^0 \gamma$, $a_{PN\Pi} = a_{\Theta p K}= -1$,
$Q_N = Q_p = 1$ and $Q_{K^0} = 0$. For $\Xi^{--}_{3/2} \to
\Sigma^- K^-\gamma$, $a_{PN\Pi} = a_{\Xi^{--}_{3/2} \Sigma^- K^-}
= -1$, $Q_P = Q_{\Xi^{--}_{3/2}} = -2$, $Q_N = Q_{\Sigma^-} = -1$,
$Q_\Pi = Q_{K^-} = -1$. And for $\Xi^{--}_{3/2} \to \Xi^-
\pi^-\gamma$, $a_{PN\Pi} = a_{\Xi^{--}_{3/2}\Xi^-\pi^-} = 1$, $Q_P
= Q_{\Xi^{--}_{3/2}} = -2$, $Q_N = Q_{\Xi^-} = -1$, $Q_\Pi =
Q_{\pi^-} = -1$.

The parameter $g_{PN\Pi}$ can be determined from a pentaquark $P$
decays into a baryon and a meson. For example

\begin{eqnarray}
{g^2_{PN\Pi}\over 2 f_\pi^2} &=& {\Gamma(\Theta^+ \to n K^+) 16\pi
m_\Theta \over (m_n + \hat P m_\Theta)^2(
(m_n -\hat P m_\Theta)^2-m^2_K)Phase},\nonumber\\
Phase &=&\sqrt{(1-(m_K+m_n)^2/m^2_\Theta)((1-(m_K -
m_n)^2/m_\Theta^2))}. \label{cpn}
\end{eqnarray}
In the above  ``$\hat P$'' is the eigenvalue of the parity, it
takes ``+'' for positive parity and ``$-$'' for negative parity
pentaquark, respectively.

From Table 1 we see that $\Theta^+$ only has two strong decay
channels, $p K^0$ and $n K^+$. The total width of $\Theta^+$ is
therefore $\Gamma_\Theta = \Gamma (\Theta^+ \to p K^0) + \Gamma
(\Theta^+ \to n K^+)$. If the $\Gamma_\Theta $ is determined, one
can determine $g^2_{PN\pi}$ from eq.\ref{cpn}

\subsection{The spin-3/2 case}

 In this
case one needs to use the Rarita-Schwinger field for pentaquarks
$P^\mu_{ilm}$. The electromagnetic couplings needed are modified
compared with spin-1/2 particles, and they are given by

\begin{eqnarray}
L_e &=& \bar P^\alpha i \gamma^\mu D_\mu P_\alpha = \bar
P^\alpha_{ijk}i\gamma^\mu (\partial_\mu
P^{ijk}_\alpha-V_{\mu,l}^iP^{ljk}_\alpha-V_{\mu,l}^jP^{ilk}_\alpha
-V_{\mu,l}^kP^{ijl}_\alpha),\nonumber\\
L_{m} &=& {\mu_P\over 4} \bar P^\alpha_{ijk}\sigma^{\mu\nu}
(f_{\mu\nu,l}^iP^{ljk}_\alpha +f_{\mu\nu,l}^jP^{ilk}_\alpha
+f_{\mu\nu,l}^kP^{ijl}_\alpha).
\end{eqnarray}

Since a spin-3/2 particle can have dipole and quadrupole moments,
if both are not zero, one should add another term to the
electromagnetic couplings,

\begin{eqnarray}
L_q = \tau_P \bar P_\nu F^{\mu\nu} P_\mu,
\end{eqnarray}
We will take it to be zero in our later discussions.

The chiral Lagrangian for strong coupling involving a pentaquark,
a baryon and a pseudoscalar is given by

\begin{eqnarray}
L_{PN\Pi} = g_{PN\Pi}\bar P^\mu_{ilm} \gamma_5\Gamma_P (A_\mu)^l_j
N^m_k \epsilon^{ijk} + H.C.
\end{eqnarray}

From the above we have

\begin{eqnarray}
\Gamma(P \to N \Pi) &=& {g^2_{PN\Pi}\over 2 f^2} {Phase\over 16
\pi m_P}{1\over 3} ((\hat P m_P +
m_N)^2 - m^2_\Pi)\nonumber\\
&\times& ({1\over 4 m^2_P}(m^2_P + m^2_\Pi - m^2_N)^2 -m^2_\Pi).
\end{eqnarray}

Combining the above information we obtain  the matrix element for
$P\to N \Pi \gamma$
\begin{eqnarray}
M( P \to  N \Pi \gamma) &=& {eg_{PN\Pi}\over \sqrt{2} f}
 a_{PN\Pi}\epsilon^*_\mu
 \bar N[ Q_{\Pi} \gamma_5\Gamma_P g^{\mu\nu} \nonumber\\
&-&(Q_N\gamma^\mu + {\mu_N\over 2} [\gamma^\mu, \gamma \cdot
P_\gamma]) {1\over \gamma\cdot P_\gamma +\gamma\cdot P_N - m_N}
\gamma_5\Gamma_P P_\pi^\nu \nonumber\\
&+&\gamma_5\Gamma_P P_\Pi^\alpha G^{\;\;\nu}_\alpha (Q_P
\gamma^\mu + {\mu_P\over 2}[\gamma^\mu,
\gamma \cdot P_\gamma])\nonumber\\
 &-& Q_\Pi {(2P_\Pi+P_\gamma)^\mu\over
(P_\Pi+P_\gamma)^2-m^2_\Pi} \gamma_5\Gamma_P ( P_\Pi +
P_\gamma)^\nu]P_\nu. \label{rs}
\end{eqnarray}

In the above $G^{\mu\nu}$ is the spin-3/2 propagator resulting
from the following most general Lagrangian\cite{11}

\begin{eqnarray}
L &=& \bar P_\mu \Lambda^{\mu\nu} P_\nu,\nonumber\\
\Lambda^{\mu\nu} &=& (\gamma\cdot P_P - m_P) g^{\mu\nu} +
A(\gamma^\mu P^\nu_P + P^\mu_P \gamma^\nu)\nonumber\\
& +& {1\over 2} (3A^2+2A+1) \gamma^\mu \gamma\cdot P_P \gamma^\nu
+ m_P (3 A^2 + 3 A + 1) \gamma^\mu\gamma^\nu.
\end{eqnarray}

The propagator is given by\cite{11}

\begin{eqnarray}
G^{\mu\nu} &=& {1\over \gamma\cdot P_P - m_P} (-g^{\mu\nu} +
{1\over 3} \gamma^\mu\gamma^\nu + {1\over 3 m_P}(\gamma^\mu
P^\nu_P - P^\mu_P \gamma^\nu) + {2\over 3 m_P^2}P_P^\mu
P_P^\nu)\nonumber\\
 &-&{1\over 3m^2_P} {A+1\over (2A+1)^2}
((2A+1)(\gamma^\mu P_P^\nu + P^\mu_P \gamma^\nu)\nonumber\\
& -&{A+1\over 2} \gamma^\mu (\gamma\cdot P_P + 2 m_P)\gamma^\nu +
m\gamma^\mu \gamma^\nu). \label{pr}
\end{eqnarray}

%

To include interaction with photon, one uses the minimal
substitution which guarantees gauge invariance to obtain the
couplings. The lowest order interaction vertex $Q_P \bar P_\alpha
\Gamma^{\alpha \beta}_\mu P_\beta$ which is different than
spin-1/2 interaction vertex $Q_P \bar P \gamma_\mu P$.
$\Gamma^{\alpha\beta}_\mu$ is given by

\begin{eqnarray}
\gamma^\mu g_{\alpha \beta} + A(\gamma_\alpha g^\mu_\beta +
g^\mu_\alpha \gamma_\beta) + {1\over 2} (3 A^2+2A+1) \gamma^\alpha
\gamma_\mu \gamma^\beta.
\end{eqnarray}

The final result is $A$ independent. In eq.\ref{rs} we have chosen
a particular case of $A= 0$ for simplicity. Therefore one should
also use $G^{\;\;\nu}_\alpha$ with $A=0$ in eq.\ref{pr}.

\section{Numerical Results}

In our numerical studies, we will concentrate on processes
involving pentaquarks with exotic quantum numbers, the $\Theta$
and $\Xi^{--}_{3/2}$. Processes involving other pentaquarks can be
similarly carried out. We now display our numerical results for
both spin-1/2 and spin-3/2, and different parities cases. For the
pentaqaurk masses, we use $m_\Theta = 1542$ MeV and $m_{\Xi_{3/2}}
= 1862$ MeV.  We will treat the magnetic dipole moments as free
parameters and let $\kappa_P = 2m_P \mu_P$ to vary between $-1$ to
$1$. The parameter $g_{PN\Pi}$ is determined by the decay width of
the pentaquark. In our calculations we will express it as a
function of $\Gamma_\theta$.

\subsection{Photoproduction}

Photoproduction of pentaquark can provide useful information about
the pentaquark properties\cite{7}. An easy way of photoproduction
of pentaqaurks is through a photon beam collides with a fixed
target containing protons and neutrons. In this case, only
production of $\Theta$ is possible via $\gamma n \to \Theta^+
K^-$, and $\gamma p \to \Theta^+ \bar K^0$. The results for the
cross sections in the laboratory frame (fixed $n$ and $p$) as
functions of photon energies for both spin-1/2 and spin-3/2 are
shown in Figs. 2 and 3.
\pagebreak[4]

\begin{figure}[!htb]
\begin{center}
\caption{Cross sections for $\gamma n \to \Theta^+ K^-$ in the
laboratory frame with spin 1/2 and 3/2. Figures a and b are for
positive and negative parities, respectively}
\begin{tabular}{cc}
{\includegraphics[width=8cm]{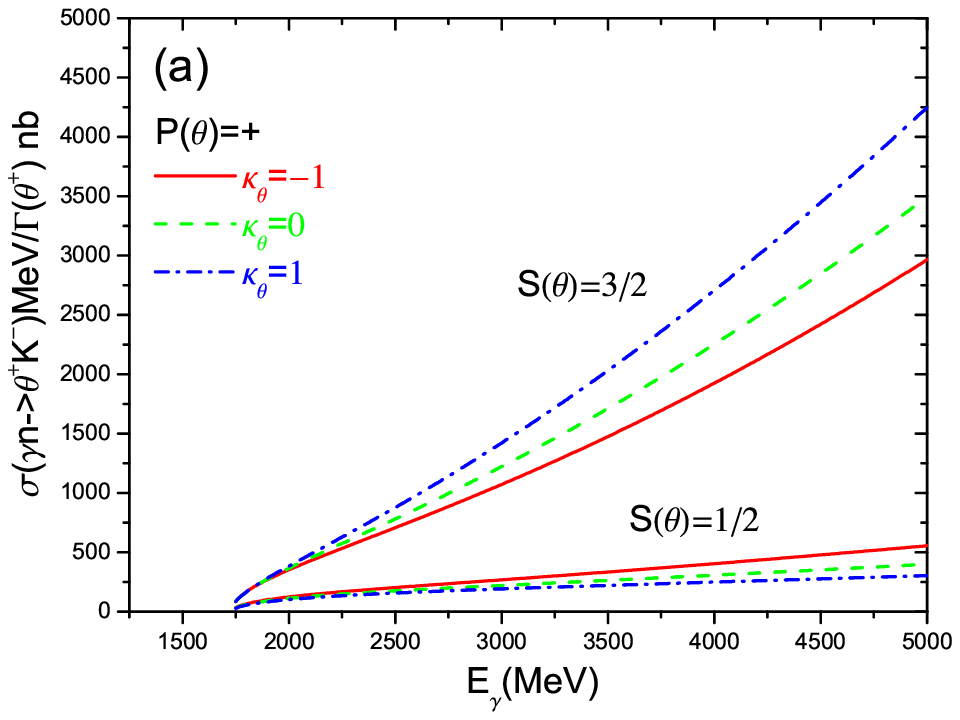}}&
{\includegraphics[width=8cm]{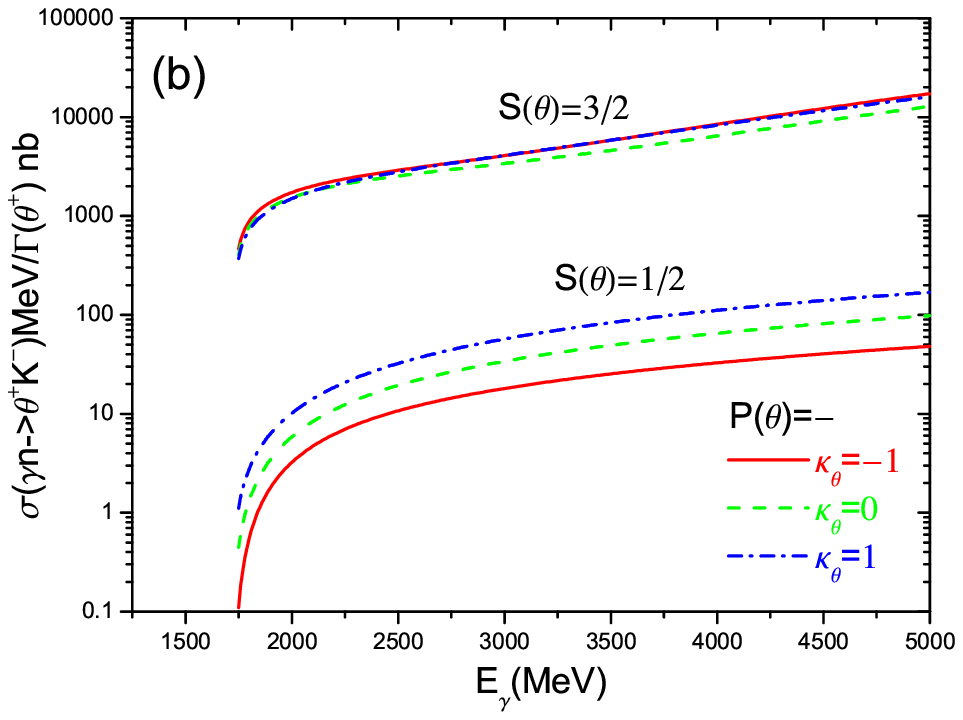}}
\end{tabular}
\end{center}
\label{fig2}
\end{figure}

\begin{figure}[!htb]
\begin{center}
\caption{Cross sections for $\gamma p \to \Theta^+ \bar K^0$ in
the laboratory frame with spin-1/2 and spin-3/2. Figures a and b
are for positive and negative parities, respectively}
\begin{tabular}{cc}
{\includegraphics[width=8cm]{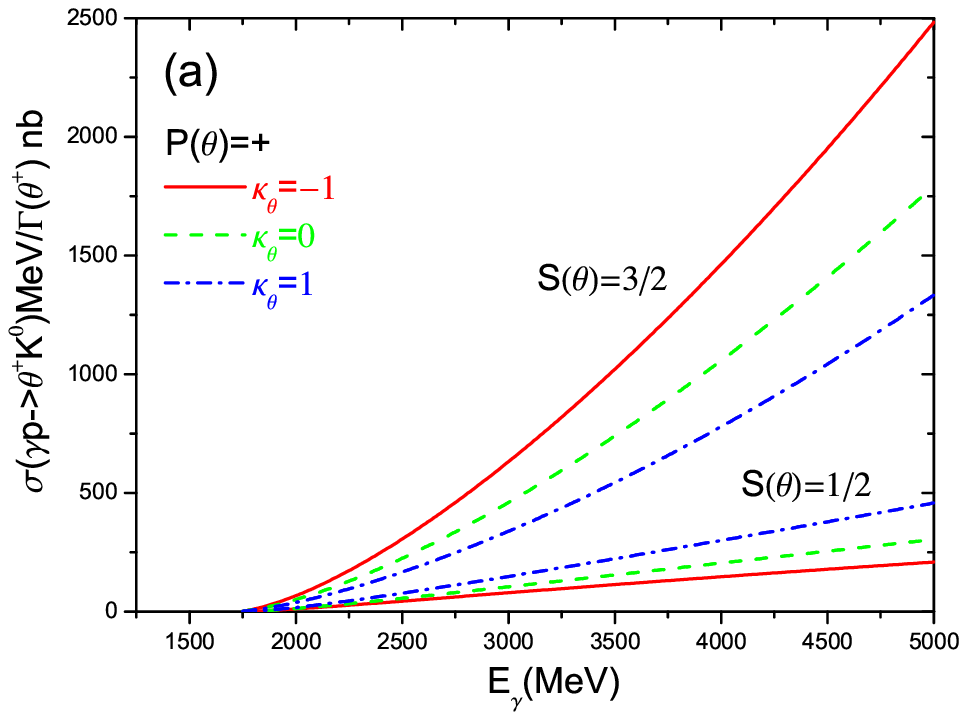}}&
{\includegraphics[width=8cm]{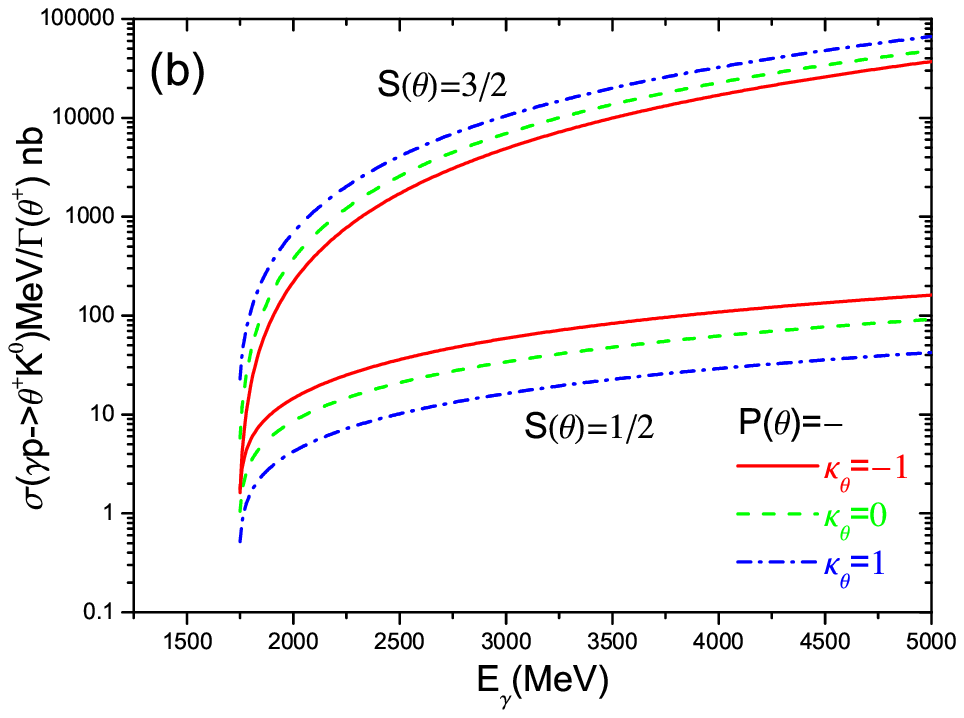}}
\end{tabular}
\end{center}
\label{fig3}
\end{figure}

From Figs 2 and 3, it can be seen that for spin-1/2 case the cross
section for $\gamma n \to \Theta^+ K^-$ with positive parity has
larger cross section than negative parity case. For example for
$\kappa_\Theta = 0$ and $E_\gamma=2.4$ GeV, the cross sections for
these two cases are $155\Gamma(\Theta^+)nb\cdot MeV^{-1}$ and
$17\Gamma(\Theta^+)nb\cdot MeV^{-1}$, respectively. The cross
section for $\gamma p \to \Theta^+ K^0$ with positive parity has
larger cross section than negative parity case, the cross sections
for these two cases are $47\Gamma(\Theta^+)nb\cdot MeV^{-1}$ and
$18\Gamma(\Theta^+)nb\cdot MeV^{-1}$, respectively.

For spin-3/2, the negative parity case has larger cross section
compared with positive parity case. For example with
$\kappa_\Theta =0$ and $E_\gamma = 2.4$ GeV, the cross sections
for $\gamma n \to \Theta^+ K^-$ are $2350\Gamma(\Theta^+)nb\cdot
MeV^{-1}$ and $691\Gamma(\Theta^+)nb\cdot MeV^{-1}$ for negative
parity and positive parity. The cross sections for $\gamma p \to
\Theta^+ K^0$ are $1953\Gamma(\Theta^+)nb\cdot MeV^{-1}$ and
$184\Gamma(\Theta^+)nb\cdot MeV^{-1}$ respectively.

One can clearly see from Figures 2 and 3 that regardless the
parity, spin-3/2 pentaquark has cross section larger than
spin-1/2. This can provide important information about the spin.
The separation between the cross sections with positive and
negative parities is large which can be used to obtain information
about the parity of the pentaqaurk too.

The cross sections also depend on magnetic dipole moment of
pentaquarks. From the figures we see that the changes in the cross
section can vary several times when $\kappa$ changes from -1 to 1.

The case for $\Theta$ with spin-1/2 has been discussed in
Ref.\cite{5,7}. Our approach is the same as that used in
Ref.\cite{5} and we agree with their results which are shown in
Fig. 2. Our approach is different than that used in Ref.\cite{7}.
This leads to the different behavior of photon energy $E_\gamma$
dependence. Detailed experimental data will provide more
information about the underlying theory for photoproduction. In
our estimate we have neglected other possible intermediate states,
such as $K^*$ which can change the cross section. But model
calculations show that $K^*$ contribution does not change the
general features\cite{7}. We expect that the results obtained here
provide a reasonable estimate.

\subsection{Radiative Decays}

Once pentaquarks are produced they  can decay radiatively through
$\Theta^+ \to \gamma K^+ n$, $\Theta^+ \to \gamma K^0 p$, and
 $\Xi^{--}_{3/2}\to \gamma K^- \Sigma^-$, $\Xi^{--}_{3/2}\to \gamma
\pi^- \Xi^-$, respectively.

It is well known that there are divergencies when photon energies
approach zero in radiative decays of the types discussed here. To
remedy these divergencies, we require that the photon energies to
be larger than $0.05$ MeV. The results for radiative $\Theta$
decays are shown in Figs. 4 and 5. The results for radiative
$\Xi^{--}_{3/2}$ decays are shown in Figs. 6 and 7.

\begin{figure}[!htb]
\begin{center}
\caption{Radiative $\Theta^+ \to \gamma n K^+$ decay for spin-1/2
and spin-3/2. Figures a and b are for positive and negative
parities, respectively}
\begin{tabular}{cc}
{\includegraphics[width=8cm]{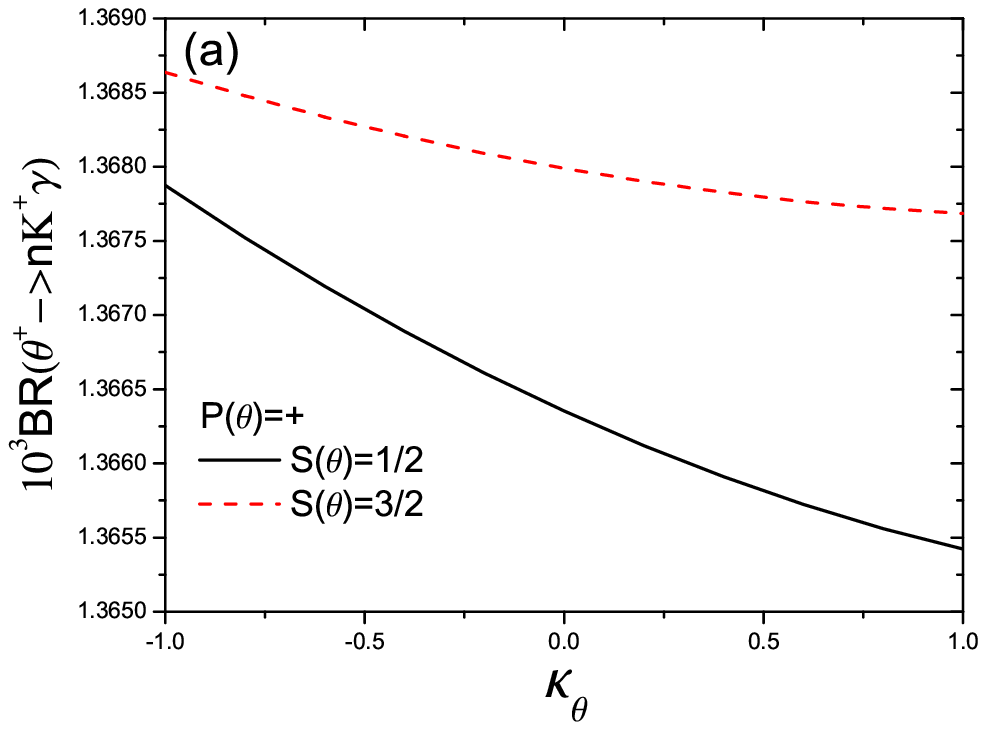}}&
{\includegraphics[width=8cm]{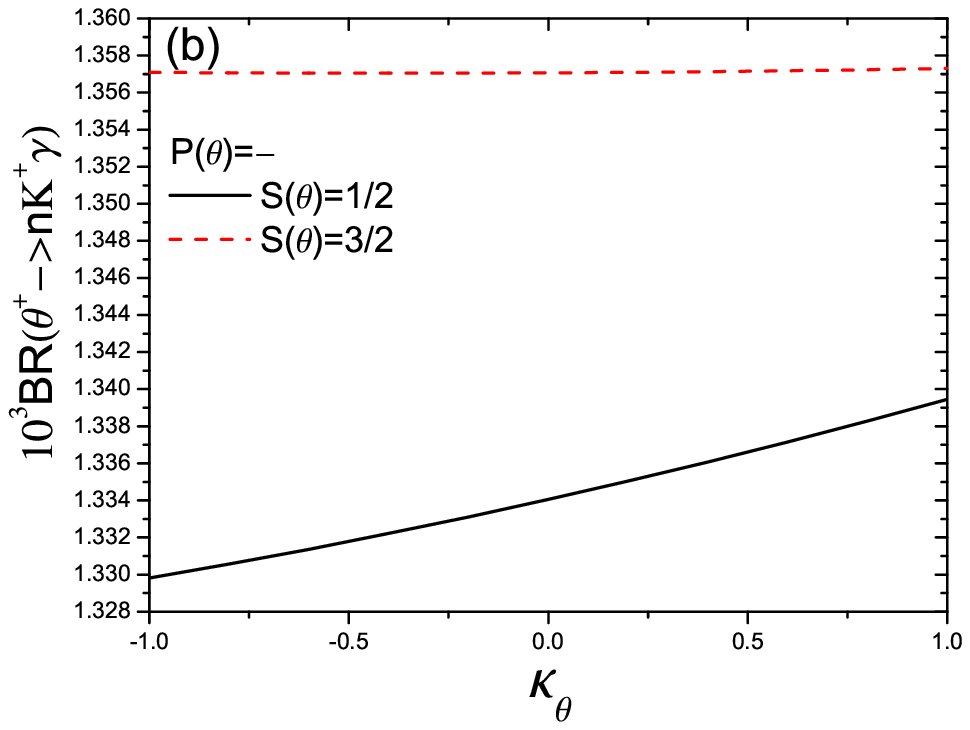}}
\end{tabular}
\end{center}
\label{fig4}
\end{figure}

\begin{figure}[!htb]
\begin{center}
\caption{Radiative $\Theta^+ \to \gamma p \bar K^0$ decay for
spin-1/2 and spin-3/2. Figures a and b are for positive and
negative parities, respectively}
\begin{tabular}{cc}
{\includegraphics[width=8cm]{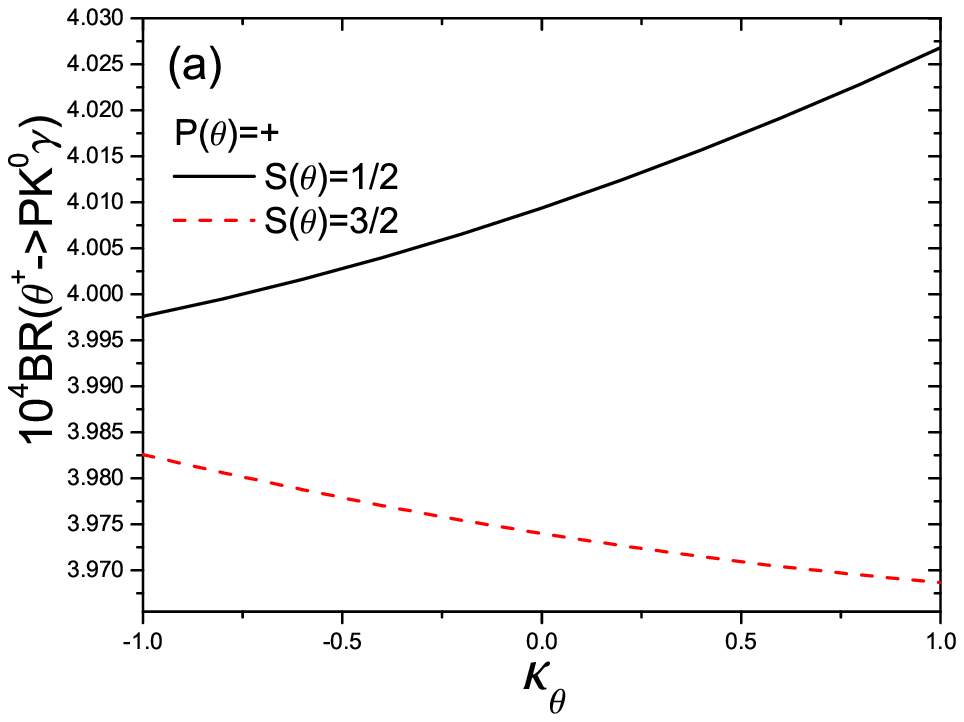}}&
{\includegraphics[width=8cm]{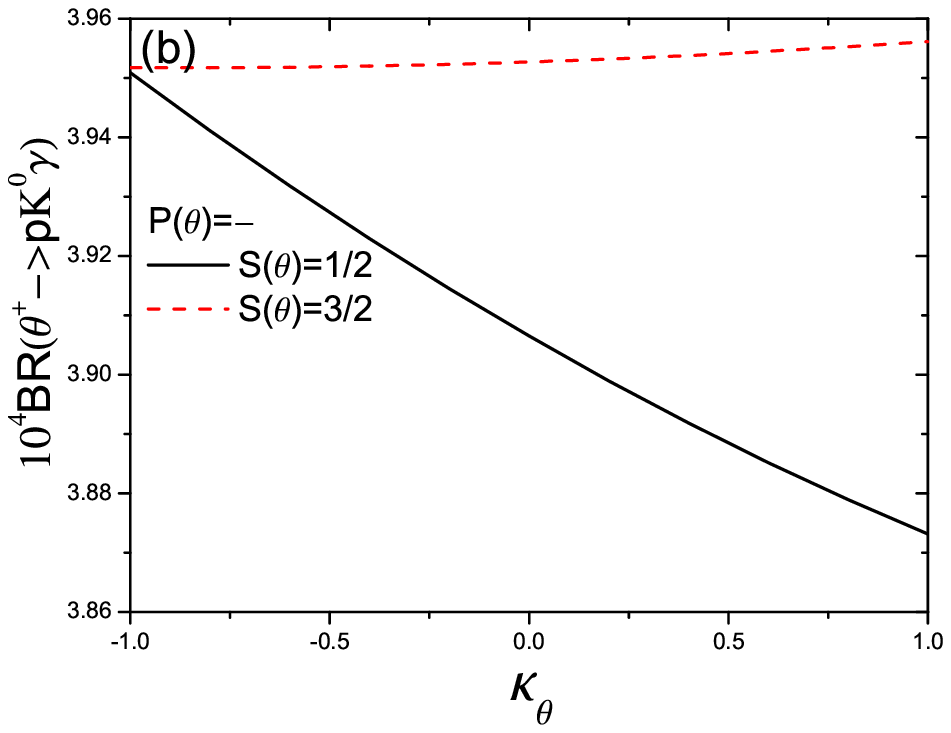}}
\end{tabular}
\end{center}
\label{fig5}
\end{figure}
\pagebreak[4]

 For $\Theta^+$ radiative decays, the branching
ratios for spin-1/2 and spin-3/2 cases are approximately
$1.3\times 10^{-3}$ and $4\times 10^{-4}$ for $\Theta^+ \to \gamma
n K^+$ and $\Theta^+ \to \gamma p K^0$, respectively. These can be
used to check the consistence of the model. However, the branching
ratios for these decays are not sensitive to the spin, parity
and anomalous magnetic dipole moment of the pentaqaurks.

The situation changes when consider radiative decays of
$\Xi^{--}$. From Figs. 6 and 7, one can see that the branching
ratios for spin-1/2 cases are about two times larger than the
branching ratios for spin-3/2 cases. It is also interesting to
note that the branching ratio for $\Xi^{--}\to \gamma \Xi^- \pi^-$
is at the level of a few percent which may be easily studied
experimentally.

\begin{figure}[!htb]
\begin{center}
\caption{Radiative $\Xi^{--}_{3/2} \to \gamma \Sigma^- K^-$ decay
for spin-1/2 and spin-3/2. Figures a and b are for positive and
negative parities, respectively}
\begin{tabular}{cc}
{\includegraphics[width=8cm]{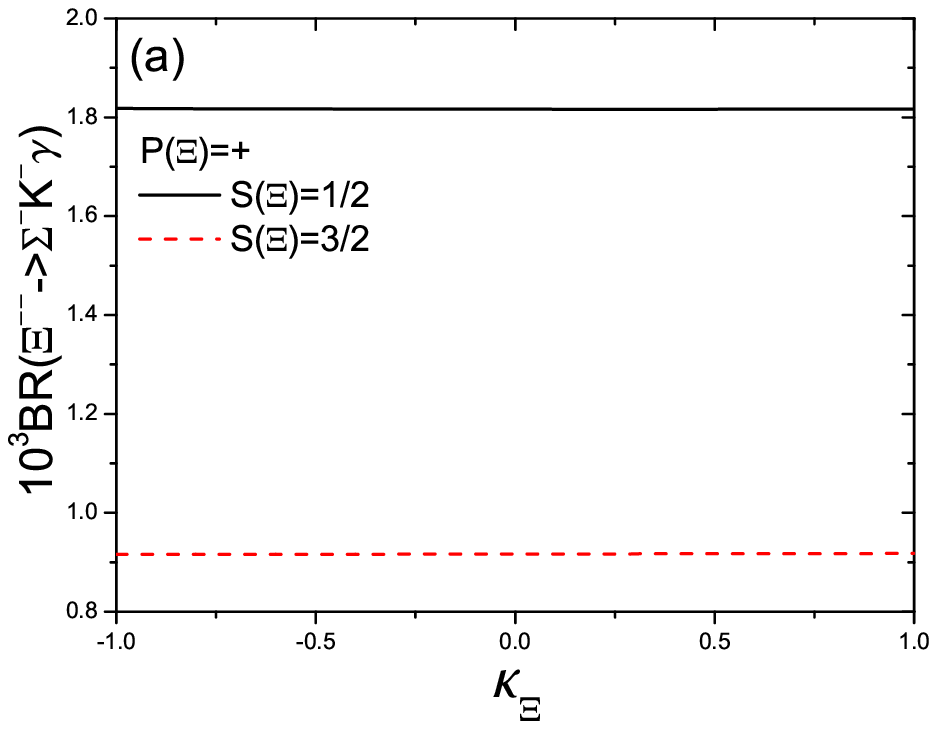}}&
{\includegraphics[width=8cm]{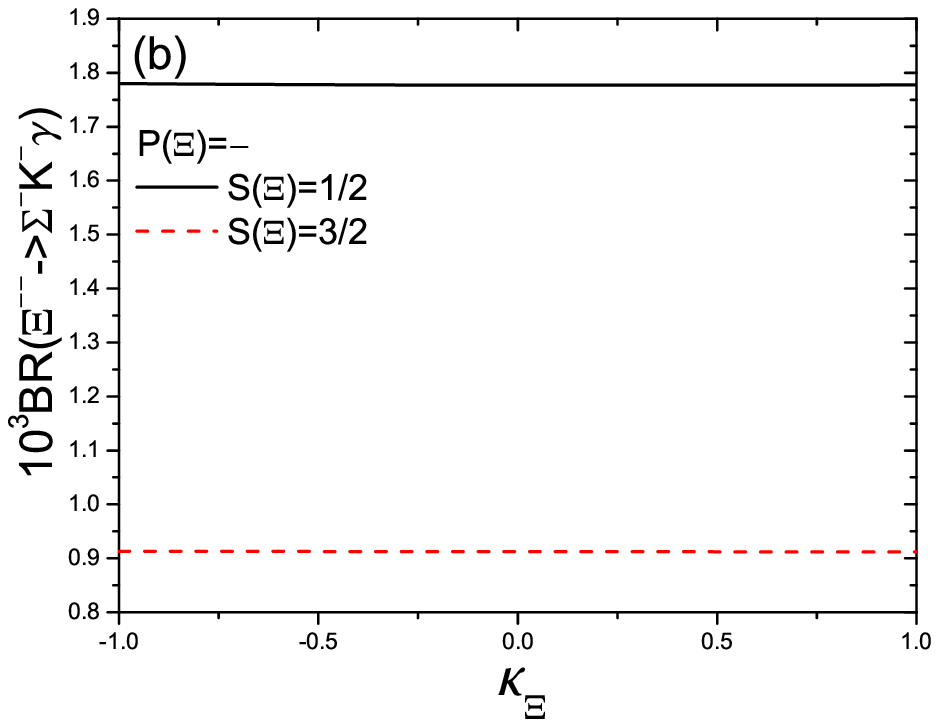}}
\end{tabular}
\end{center}
\label{fig6}
\end{figure}

\begin{figure}[!htb]
\begin{center}
\caption{Radiative $\Xi^{--}_{3/2} \to \gamma \Xi^- \pi^-$ decay
for spin-1/2 and spin-3/2. Figures a and b are for positive and
negative parities, respectively}
\begin{tabular}{cc}
{\includegraphics[width=8cm]{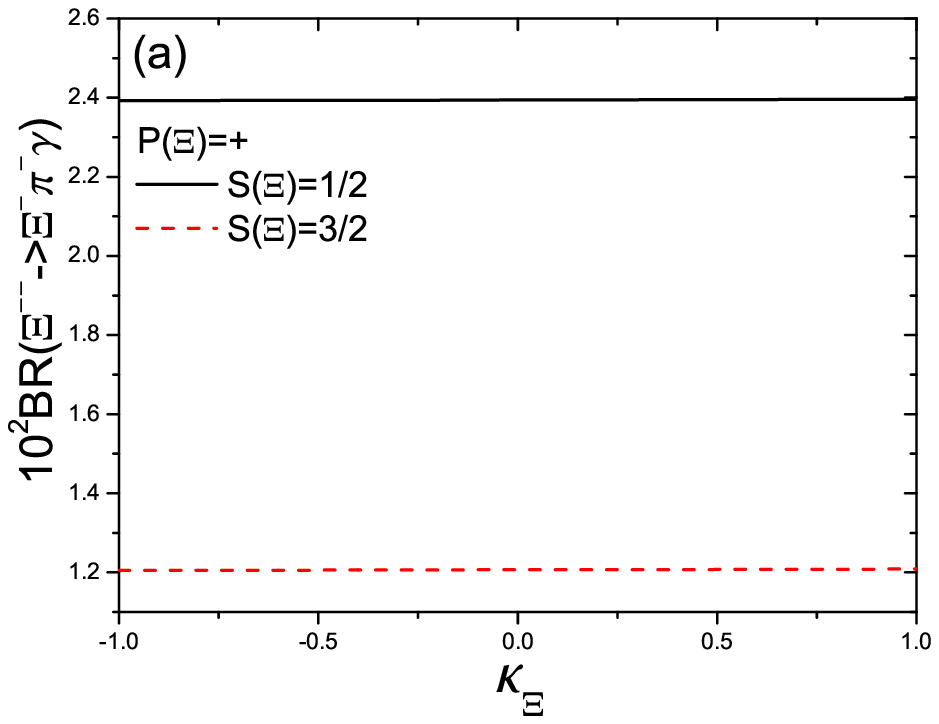}}&
{\includegraphics[width=8cm]{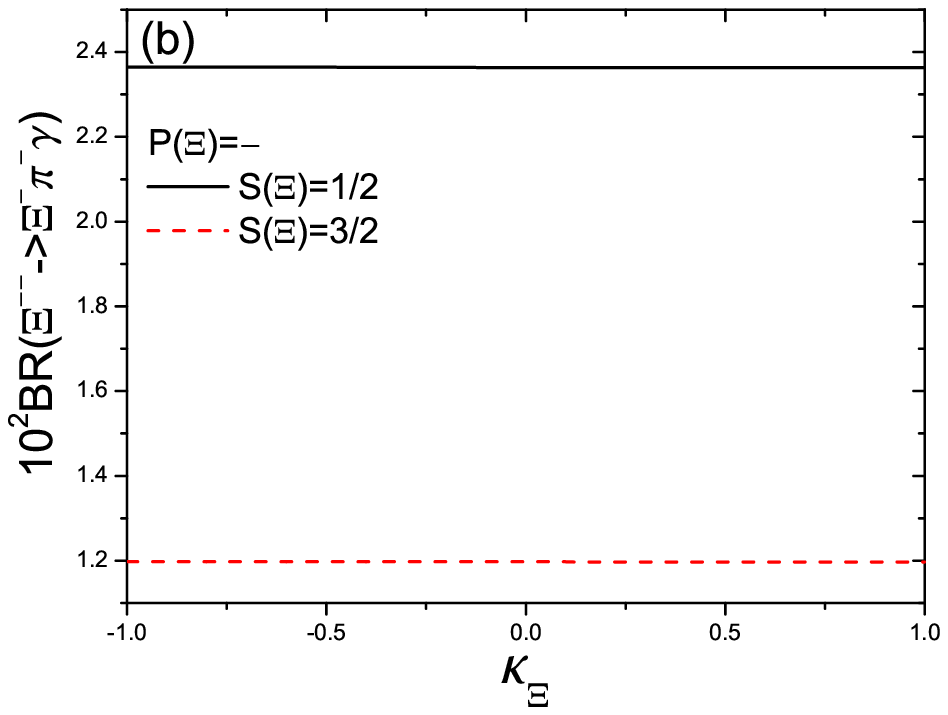}}
\end{tabular}
\end{center}
\label{fig7}
\end{figure}

In conclusion we have studied several radiative processes of
pentaquarks using chiral perturbatin theory. We find that the
photoproduction cross sections of $\Theta^+$ are sensitive to the
spin, parity and anomalous magnetic dipole moment of the
pentaquark. Radiative decays of $\Theta^+$ can also provide
consistent check of the theory although these decays are not very
sensitive to the spin, parity and anomalous magnetic dipole
moment. Radiative decays of $\Xi^{--}$ are sensitive to the spin
of the pentaquark. Near future experiments on pentaquark radiative
processes can provide important information about pentaquark
properties.

\end{document}